\documentstyle{article}

\begin{document}

\headsep 1cm

\title{
Time asymmetries in quantum cosmology and the searching for
boundary conditions to the Wheeler-DeWitt equation \vskip1cm}

\author{\noindent
Mario Castagnino
\thanks{castagni@iafe.uba.ar}, Gabriel Catren
\thanks{catren@iafe.uba.ar}
\ and Rafael Ferraro
\thanks{ferraro@iafe.uba.ar}
\\ \\ 
Instituto de Astronom\'\i a y F\'\i sica del Espacio,\\ Casilla de
Correo 67, Sucursal 28, 1428 Buenos
Aires, Argentina\\ \\ 
Departamento de F\'\i sica, Facultad de Ciencias Exactas y
Naturales,\\ Universidad de Buenos Aires, Ciudad Universitaria,
Pabell\' on I,\\ 1428 Buenos Aires, Argentina}

\date{}

\maketitle

\vskip1cm

\begin{abstract}
The paper addresses the quantization of minisuperspace
cosmological models by studying a possible solution to the problem
of time and time asymmetries in quantum cosmology. Since General
Relativity does not have a privileged time variable of the
newtonian type, it is necessary, in order to have a dynamical
evolution, to select a physical clock. This choice yields, in the
proposed approach, to the breaking of the so called clock-reversal
invariance of the theory which is clearly distinguished from the
well known motion-reversal invariance of both classical and
quantum mechanics. In the light of this new perspective, the
problem of imposing proper boundary conditions on the space of
solutions of the Wheeler-DeWitt equation is reformulated. The
symmetry-breaking formalism of previous papers is analyzed and a
clarification of it is proposed in order to satisfy the
requirements of the new interpretation.
 \vskip 1cm

\end{abstract}

\vskip  1cm

\newpage

\section{Introduction}

\smallskip The so called `problem of time' in quantum gravity is one of
the main conceptual and technical problems of the canonical
quantization program
(Ref.\cite{Kuchar},\cite{isham},\cite{ferraro}). This quantization
program begins by reformulating General Relativity under a
Hamiltonian formulation (ADM formalism, see \cite{gravitation}).
Within the framework of this formalism the Lorentzian space-time
manifolds $M$ are split in a collection of spacelike hypersurfaces
$\sum$ parametrized by a real time parameter $t$ (foliation of the
space-time). The role of the canonical variables is played by the
Riemannian metrics $g_{ij}$ of these hypersurfaces $\sum.$ The
corresponding configuration space is the space of all the possible
Riemannian metrics $g_{ij}$ which is called \emph{superspace}. The
conjugated momenta $\pi ^{ij}$ are related to the extrinsic
curvature of the hypersurfaces $\sum$, i.e., to the way in which
these hypersurfaces are embedded in the space-time manifold $M$.
In this Hamiltonian formulation the general covariance of the
theory appears as a set of constraints among the canonical
variables (four constraints per each point of space-time). The
so-called Hamiltonian constraint assures the invariance of the
theory under changes of the foliation of the space-time. The
momenta constraints (three per each point of space-time) assure
the invariance of the theory under a change of the spatial
coordinates used to represent the spatial geometry of each
hypersurface $\sum $. The existence of the Hamiltonian constraint
means that the theory does not select a privileged time variable.
To consider General Relativity as a dynamical system it is thus
necessary to choose a physical clock, i.e., a physical degree of
freedom with suitable properties, to play the role of time. On the
contrary, in quantum mechanics there is a privileged time variable
which is an evolution parameter clearly separated from the other
degrees of freedom which are associated with quantum operators.
This difference between General Relativity and quantum mechanics
is the main problem for finding a quantum theory of gravity in the
framework of the canonical approach.

In this paper the choice of a physical clock will be treated as a
symmetry breaking of a certain kind of time symmetry
(clock-reversal transformations) which is induced by the
double-sheet structure of the Hamiltonian constraint. It will be
addressed the case of a time independent reduced Hamiltonian,
leaving for another work the treatment of time dependent
Hamiltonians. In section II the parametrized system formalism is
reviewed and its conceptual problems are discussed. A
reconceptualization of this formalism is proposed in order to
solve these problems. In section III an example of a reducible
minisuperspace model is presented. In section IV the fundamental
distinction between clock-reversal transformations and
motion-reversal transformations is proposed for time independent
reduced Hamiltonians in order to solve an apparent paradox of the
worked approach. Its classical and quantum versions are defined.
In section V the consequences of our approach for the problem of
the boundary conditions on the space of solutions of the
Wheeler-DeWitt equation are studied. In section VI a general
formalism to study irreversible processes via a symmetry breaking
framework is reviewed and its adaptation for the case of General
Relativity is presented in order to fix the proposed approach in a
mathematical framework.

\section{Parametrized systems formalism}

One of the main properties of the Hamiltonian structure of General
Relativity is the presence of the Hamiltonian constraint $H=0$. As
stated above this constraint means that the action does not depend
on the foliation chosen to describe the ``evolution'' in the
Hamiltonian formulation, this invariance being a consequence of
the covariance of the theory. A well known formalism which has
this kind of invariance is the parametrized system formalism
(Ref.\cite{Kuchar}, \cite{lanczos}). In a parametrized system the
absolute time $t$ is added to the dynamical variables leaving this
increased set of dynamical variables as functions of an
\emph{ad-hoc} introduced irrelevant parameter $\tau $. This kind
of systems are frequently used as a paradigm for understanding the
Hamiltonian structure of General Relativity. Let us start with an
action of the form
\[
S\left[ q^{\mu },p_{\mu }\right] =\int_{t_{1}}^{t_{2}}p_{\mu
}dq^{\mu }-h\left( q^{\mu },p_{\mu }\right) dt,\ \ \mu =1,...,n
\]
The original set of dynamical variables $\left\{ q_{\mu },p^{\mu
}\right\} $ $\left( \mu =1,...,n\right) $ is extended by
identifying $q^{0}\equiv t,$ $ p_{0}\equiv -h$. The new set of
variables are left as functions of an irrelevant parameter $\tau
$. \ The extended set $\left\{ q^{0},q^{\mu },p_{0},p_{\mu
}\right\} $ can be varied independently, provided that the
definition of $p_{0}$ is incorporated into the action as a
constraint
\begin{equation}
H=p_{0}+h=0  \label{po}
\end{equation}
with the corresponding Lagrange multiplier $N$ yielding the
following action
\begin{equation}
S\left[ q^{i}\left( \tau \right) ,p_{i}\left( \tau \right)
,N\left( \tau \right) \right] =\int_{t_{1}}^{t_{2}}\left(
p_{i}\frac{dq^{i}}{d\tau } -NH\right) d\tau  \label{pl}
\end{equation}
The presence of the Lagrange multiplier $N\left( \tau \right) $
means that the dynamics remains ambiguous in the irrelevant
parameter $\tau $ (one could say that it has no sense to speak
about dynamics until the hidden time is recovered). A constraint
like this can be disguised by scaling it with a function $f\left(
q,p\right) $ of a definite sign on the constraint surface or by
performing a canonical transformation
\[
\left\{ q^{i},p_{i}\right\} =\left\{ q^{o}=t,p_{0}=-h,q^{\mu
},p_{\mu }\right\} \rightarrow \left\{ Q^{i},P_{i}\right\}
\]
where now the time is hidden among the rest of the dynamical
variables.

A main argument against this interpretation of the Hamiltonian
structure of General Relativity (Ref.\cite{barbour}) is that its
Hamiltonian constraint is quadratic in all its momenta while the
Hamiltonian constraint of a parametrized action (\ref{po}) is
linear in the momentum conjugated to time. In fact the
parametrized system formalism is not completely fitted to describe
General Relativity because this formalism supposes that an
absolute external time is hidden among the dynamical variables. In
this approach to reduce the system means to find the hidden time
by performing canonical transformations or scaling the Hamiltonian
constraint in order to find a constraint linear in the momentum
canonically conjugated to time. We think that the existence of
this hidden time is an unfounded supposition based on an
extrapolation of our experience in other branches of physics in
which there is always an external time parameter. The most
singular feature of modern Cosmology is that it studies a system
which, by definition, does not live in an external scenario. We
think that the corresponding physics of this kind of
auto-contained systems must depart in many fundamental ways from
usual physical theories. In particular, the supposition that there
is a privileged time hidden among the canonical variables
represents a tentative to reduce General Relativity to the usual
pattern of what a physical theory is supposed to be. We consider
that one of the most fundamental properties of General Relativity
is that its solutions does not represent, in general, a time
evolution of certain dynamical variables, but that it is a theory
which selects certain relative (not dynamical) configurations of
its canonical variables which, under certain conditions, could be
considered as dynamical evolutions if proper \emph{physical
clocks} can be selected. It is thus only possible to speak about
physical clocks, i.e., degrees of freedom which can play the role
of evolution parameters for the others degrees of freedom. The
only requirement to be satisfied by a degree of freedom in order
to be a proper physical clock is that it should be possible to
express any relative configuration of the n canonical variables as
n-1 \emph{functions} of the variable $q_{0}$ chosen as the
physical clock. As many authors have pointed out
(Ref.\cite{barbour},\cite{rovelli}) we can never observe the
evolution along the newtonian time flow like $q_{1}\left( t\right)
$ and $q_{2}\left( t\right) $ but rather the evolution of certain
variables relative to the change of another variable, i.e.,
something like $q_{2}\left( q_{1}\right) $\footnote{In 1918 the
philosopher Ludwig Wittgenstein wrote in the 6.3611 proposition of
his \emph{\ Tractatus Logico-philosophicus}: ``We can no compare
any process with `the flow of time' -which does not exist-, but
with another process (as the motion of a chronometer, for
example). Therefore the description of the flow of time is only
possible using another process.''}. In this relational approach we
cannot say that reducing the system means to find the hidden time
but that reducing the system means to select, among the canonical
variables, a proper physical clock. A main consequence of this
subtle and fundamental change in the perspective is that, as there
is not a privileged time, all the momenta must appear
quadratically in the Hamiltonian constraint, i.e., all the momenta
must appear on an equal footing (as effectively happens in General
Relativity), being this an essential fact of the theory which
turns the parametrized system formalism an improper analogy. It is
thus necessary to reformulate the model which is intended to mimic
General Relativity, in order to properly describe this substantial
difference. This reformulation must accomplish the requirement
that all the canonical momenta must appear quadratically in the
Hamiltonian constraint in order not to privilege a certain clock
among others. There is even another and more important reason
which, if we follow this new interpretative framework, turns
essential the fact that the Hamiltonian constraint must be
quadratic in all its momenta. If there is not a privileged time
the solutions are necessarily statics trajectories, i.e., relative
configurations among the different variables, for example
$q_{2}\left(
q_{1}\right) $. If one wants now to select a physical clock, for example $%
q_{1}$(we are supposing that $q_{1}$ is a monotonic function along
the trajectory) there is still an ambiguity, i.e., one still have
to choose in which direction the trajectory is being unfold. This
means that one can choose $t=q_{1}$ or $t=-q_{1}.$ The static
trajectory does not privilege any direction and so both kind of
solutions must appear in the reduced formalism. We will show that,
for reducing the system, one has to separate the Hamiltonian
constraint in two sheets corresponding each sheet to each choice
of the direction in which the trajectory is unfold. In order to
make this factorization the Hamiltonian constraint must be
quadratic in the momentum conjugated to $q_{1}.$ This is the main
difference between our approach and the parametrized system
formalism. In this last framework the real time was certainly
hidden with its direction of evolution, being thus unnecessary the
presence of the other sheet. If, on the contrary, one begins with
an static configuration both directions must appear. In this new
light the Hamiltonian constraint of General Relativity not only
implies that the theory is invariant under a change of the chosen
foliation of space-time but also that it is invariant under an
inversion in the direction in which the corresponding
hypersurfaces of simultaneity are unfold.

\smallskip We will then suppose that the Hamiltonian of the model under
study can be taken (using a suitable canonical transformation
and/or scaling the Hamiltonian) to the form
\begin{equation}
H=\left( p_0+h\right) \left( p_0-h\right) =p_0^2-h^2\left( q^\mu
,p_\mu \right)  \label{poi}
\end{equation}

\smallskip Imposing the constraint $H=0$ is equivalent to choose a sheet of
the constraint surface, i.e., to select a direction for the
variable $q^{0}$ which will play the role of time in the reduced
formalism. The action (\ref{pl}) expressed in these new variables
is
\[
S\left[ q^{i}\left( \tau \right) ,p_{i}\left( \tau \right)
,N\left( \tau \right) \right] =\int p_{\mu }dq^{\mu
}+p_{0}dq^{0}-N\left( p_{0}+h\right) \left( p_{0}-h\right) d\tau
\]

The Hamilton equation for $q^0$ is
\[
\frac{dq^0}{d\tau }=N\frac{\partial H}{\partial p_0}=2Np_0
\]

As $h>0$ then $p_{0}$ never vanishes on the constraint surface.
This implies that $q^{0}$ can be made a monotonous function of
$\tau$ along each trajectory by a proper gauge choice. In this way
$q^{0}$ acquires the rank of an internal clock. In fact, we will
choose the direction of time $t$ as the increasing direction of
the variable $q^{0}$. We can do it by means of the gauge fixing condition $%
t\equiv q^{0}=\tau $, which is equivalent to choose the Lagrange multiplier $%
N:$%
\[
\frac{dq^{0}}{d\tau }=\frac{dt}{d\tau }=2Np_{0}=1
\]
or
\[
N\left( \tau \right) =\frac{1}{2p_{0}\left( \tau \right) }
\]
Then the action takes the form
\[
S\left[ q^{i}\left( q^{0}\right) ,p_{i}\left( q^{0}\right) \right]
=\int p_{\mu }dq^{\mu }+p_{0}dq^{0}-\frac{1}{2p_{0}}\left(
p_{0}+h\right) \left( p_{0}-h\right) dq^{0}
\]

In order to finish the reduction process one has to deduce in
which sheet of the constraint surface one is working: $p_0+h=0$ or
$p_0-h=0.$ The chosen
gauge fixing condition $t\equiv q^0=\tau $ means that the chosen sheet is $%
p_0+h=0.$ The other sheet does not yield the action to the
standard form of a non parametrized system
\begin{equation}
S\left[ q^\mu \left( q^0\right) ,p_\mu \left( q^0\right) \right]
=\int p_\mu dq^\mu -h\left( q^\mu ,p_\mu \right) dt  \label{ac}
\end{equation}
with a positive reduced Hamiltonian $h\left( q^\mu ,p_\mu \right)
>0$.

The constraint $p_0+h=0$ means that the chosen $N$ is
\[
N=-\frac 1{2h}
\]

If one had chosen the decreasing direction of $q^0$ as the
increasing direction of time, i.e., $t\equiv -q^0=\tau ,$ the
gauge fixing condition would have been
\[
\frac{dq^0}{d\tau }=-\frac{dt}{d\tau }=2Np_0=-1
\]

The condition $\frac{dq^0}{d\tau }=-1$ means now that the Lagrange
multiplier is
\[
N=-\frac 1{2p_0}
\]

In order to take the system to the standard reduced form
(\ref{ac}) with a positive reduced Hamiltonian\ \ $h\left( q^\mu
,p_\mu \right) >0$ one has to work on the sheet $p_0-h=0.$ This
has as a consequence that $N$, as a function of the reduced
variables $\left( q^\mu ,p_\mu \right) $, is still
\[
N=-\frac 1{2h}
\]
and one reobtains (\ref{ac}).

\section{Minisuperspace example}

In the literature about minisuperspace models it can be found many
examples of reducible models, i.e., cosmological models where a
physical clock can be separated from the rest of the dynamical
variables (Ref. \cite{simeone}). These models can be classified in
those where time is only a function of the configuration variables
(intrinsic time) and those where time is a function of the phase
space variables (extrinsic time). The Friedmann-Robertson-Walker
universe for $k=0,-1$ with cosmological constant coupled with a
massless scalar field and the Kantowski-Sachs model are examples
of the first kind (with time dependent reduced Hamiltonians). The
Taub model is a particularly interesting case because it does not
have an intrinsic time but can be reduced by an extrinsic time
with a time independent reduced Hamiltonian. The Taub model
represents an homogeneous but anisotropic universe. The
corresponding configuration space (minisuperspace) is a two
dimensional manifold parametrized by a parameter $\beta _{+}$\
measuring the spatial anisotropy and a parameter $\Omega$\
measuring the volume of the Universe. The Hamiltonian constraint
for this model is
\begin{equation}
H=-p_\Omega ^2+p_{+}^2+12\pi ^2e^{-4\Omega }(e^{-8\beta
_{+}}-4e^{-2\beta _{+}})
\end{equation}
while the momenta constraint are identically satisfied. This
constraint does not have a positive potential an it is thus not
possible to appreciate the double sheet Hamiltonian structure of
the constraint surface. The reduction of the Taub universe was
studied in Ref. \cite{taub}. By means of the coordinate
transformation
\begin{eqnarray*}
\Omega &=&v-2u \\ \beta _{+} &=&u-2v
\end{eqnarray*}
the Hamiltonian constraint can be written as
\begin{equation}
H=\frac 16\left( p_v^2+36\pi ^2e^{12v}\right) -\frac 16\left(
p_u^2+144\pi ^2e^{6u}\right)
\end{equation}
Performing the canonical transformation
\begin{eqnarray*}
q &=&Arc\sinh \left( -\frac{p_v}{6\pi }e^{-6v}\right) \\ p_q^2
&=&\frac 1{36}\left( p_v^2+36\pi ^2e^{12v}\right)
\end{eqnarray*}
whose generating function is
\[
F_1\left( v,q\right) =-\pi e^{6v}\sinh q
\]
it is possible to take the constraint to the form
\begin{equation}
H=6p_q^2-\frac 16\left( p_u^2+144\pi ^2e^{6u}\right)
\end{equation}
In this way a physical clock $q$ was separated with a reduced
Hamiltonian $h$ which does not depend on time $q$. It is now
necessary to choose a direction of $q$ for the increasing
direction of time. The last expression can be factorized in the
form of (\ref{poi})
\begin{equation}
H=\left( \sqrt{6}p_q+\frac 1{\sqrt{6}}\sqrt{p_u^2+\pi
^2e^{6u}}\right) \left( \sqrt{6}p_q-\frac
1{\sqrt{6}}\sqrt{p_u^2+144\pi ^2e^{6u}}\right)
\end{equation}

The constraint $H=0$\ is fulfilled if one of the factors vanishes
on the constraint surface. To choose which factor is null is
equivalent to choose which direction of $q$ is the increasing
direction of time. The other factor has, on the constraint
surface, a definite sign, so being possible to rescale the
Hamiltonian by this factor. In Ref. \cite{taub} the increasing
direction of $q$ was selected as time, i.e., $q=t$.

\section{Clock-reversal and motion-reversal transformations}

In some sense one could say that each choice ($t=q$ or $t=-q$)
corresponds to a kind of time reversal of the other one. If this
were the case the choice of the direction of time would be like a
breaking of the time-reversal symmetry of the original theory. But
one knows that each sheet of the Hamiltonian constraint contains a
classical system with the well known classical and quantum
symmetries under time reversals. This point is subtle and deserves
special attention in order to circumvent this apparent paradox.
Classical mechanics is a theory which is said to be invariant
under time reversals. By this one means that, given a classical
trajectory $\left\{ q\left( q_{0},p_{0},t_{0},t\right) ,p\left(
q_{0},p_{0},t_{0},t\right) \right\} $ which unfolds between $\left\{ q_{0},p_{0}\right\}$ at time $t_{0}$ to $\left\{ q_{f},p_{f}\right\}$ at time $%
t_{f}$, there exists another trajectory which seems to be the time
reversal of the former, and which is also a solution of the
Hamilton equations. This inverted trajectory is
\begin{eqnarray}
q^{mr}\left( q_{0}^{mr}=q_{f},p_{0}^{mr}=-p_{f},t_{0,}t\right)
=q\left( q_{f},-p_{f},t_{0,}t\right)  \label{oi} \\ p^{mr}\left(
q_{0}^{mr}=q_{f},p_{0}^{mr}=-p_{f},t_{0,}t\right) =p\left(
q_{f},-p_{f},t_{0,}t\right) \nonumber
\end{eqnarray}
and exists provided that the Hamiltonian is quadratic in $p$ and
does not depend on $t$ (the meaning of the superindex $^{mr}$ will
be explained below). It is often said that the operation of
passing from a certain trajectory $\left\{ q\left( t\right)
,p\left( t\right) \right\} $ to the one defined by (\ref{oi}) is
like ``playing the film backwards''. Actually this assertion does
not do enough justice to the solution (\ref{oi}) because it
darkens the role of the clock: if the movie is played backwards
one would see also the hands of the clock running backwards. Of
course the solution (\ref{oi}) refers to a clock going forward,
but with initial conditions which have been inverted with respect
to the original trajectory: the new trajectory starts with an
inverted velocity from the point where the original one ends, but
it starts at the same time than the original one and unfolds in
the same direction of time. We will call the operation (\ref{oi})
a \emph{motion-reversal transformation} (this is the reason for
the superindex $mr$ in (\ref{oi})). It is a remarkable fact that
the double sheet Hamiltonian constraint surface induces a
different kind of time symmetry: passing from one sheet to the
other one is equivalent to the change $t\rightarrow -t$,
$h\rightarrow -h$. We reserve the name of \emph{clock-reversal
transformation} for this second kind of time symmetry. The
motion-reversal transformation represents a motion with the
direction of unfolding of all the canonical variables inverted but
the one used as a physical clock, while the clock-reversal
transformation represents a motion with the evolution of all the
variables inverted including the one representing the physical
clock. Summarizing, each solution has its corresponding
motion-reversed solution on the same sheet and both of these
motions are connected by a clock-reversal transformations with a
companion pair on the other sheet. In order to fix ideas let us
suppose a dynamical system composed of two variables $q_{1}$ and
$q_{2}$ with a Hamiltonian constraint
$H(q_{1},q_{2},p_{1},p_{2})=0.$ Without loss of generality let us
suppose that, in a particular solution, the representative
configuration
point $\left( q_{1},q_{2}\right) $ makes a motion passing by $%
\left( q_{1}=-1,q_{2}=A\right) $ and $\left(
q_{1}=1,q_{2}=B\right) $. As we still did not choose a physical
clock this is not really a motion but a static trajectory. Let us
suppose that $q_{1}$ behaves as a physical clock, i.e., that there
is no two values of $q_{2}$ for the same $q_{1}$. As was said
before one has two options: $t=q_{1}$ or $t=-q_{1}$. Let us
suppose that the increasing direction of $q_{1}$ is chosen as
time, i.e., that $t=q_{1}$. It is only now that one can say that
the dynamical variable $q_{2}$ is moving from $A$ to $B$ as the
time $t$ $\left( =q_{1}\right) $ flows (figure 1(a)). From this
``original solution'' one can construct three others solutions
which corresponds to the motion-reversal of the original one
(figure 1(b)), the clock-reversal of the original one (figure
1(c)) and the clock-reversal of the motion-reversal of the
original one (figure 1(d)). In fact one could find the so called
motion-reversal trajectory of the original solution defined in
$\left( \ref {oi}\right) $. This trajectory goes from $\left(
q_{1}=-1,q_{2}=B\right) $ to $\left( q_{1}=1,q_{2}=A\right) $. In
the configuration space $\left( q_{1},q_{2}\right) $ this is
another trajectory which solves the Hamilton equations and for
which time $t$ is still increasing in the direction of the
increasing $q_{1}$, i.e., the dynamical variable $q_{2}$ moves now
from $B$ to $A$ as the time $t$ $\left( =q_{1}\right) $ flows
(figure 1(b)). Let us suppose now
that we choose the decreasing direction of $q_{1}$ as time, i.e., that $%
t\equiv -q_{1}$. This choice lead us to the clock-reversals of the
former solutions. For example the figure 1(c) represents the
clock-reversal of 1(a). The dynamical variable $q_{2}$ moves now
from $B$ to $A$ as the time $t$ $\left( =-q_{1}\right) $ passes.
It is now that the representative point is traveling the original
trajectory in the opposed direction. The motion-reversal $\left(
\ref{oi}\right) $ of this last solution is equivalent to the
clock-reversal of the motion-reversal of the original solution
(figure 1(d) is the motion-reversal of 1(c) and the clock-reversal
of 1(b)). This example should clarify the difference between the
motion-reversal operations $\left( \ref{oi}\right) $ used in
classical and quantum mechanics and the passage from one sheet of
the Hamiltonian constraint to the other one (clock-reversal
operations). The confusion between this two operations is rooted
in the fact that we are used to think the problem in the reduced
configuration space which is like looking at films in which the
motions of the hands of the clock have not been recorded. In the
reduced configuration space (the axis $q_{2}$ in the example
above) these four related solutions reduces to two and this
substantial difference degenerates.

\subsection{Classical transformations}

The Hamilton equations for the reduced variables are
\begin{eqnarray*}
\frac{dq^\mu }{d\tau } &=&N\frac{\partial H}{\partial p_\mu } \\
\frac{dp_\mu }{d\tau } &=&-N\frac{\partial H}{\partial q^\mu }
\end{eqnarray*}
Choosing $q$ as time, i.e., fixing $t\equiv q=\tau $ the Hamilton
equations take the form
\begin{eqnarray}
\frac{dq^\mu }{dq} &=&\frac{\partial h}{\partial p_\mu }
\label{pu} \\ \frac{dp_\mu }{dq} &=&-\frac{\partial h}{\partial
q^\mu }  \nonumber
\end{eqnarray}
As it was said before it is known that, given a certain trajectory
 the motion-reversal trajectory
$\left( \ref{oi}\right) $ is also a solution of the Hamilton
equations of motion. We will now define the clock-reversal
solution $\left\{ q^{\mu ^{cr}},p_\mu ^{cr}\right\} $\ by noting
that it is equal to the motion-reversal one plus an inversion of
the physical clock $t\rightarrow t^{cr}=-t$
\begin{eqnarray}
q^{\mu ^{cr}}\left( q_0^{\mu ^{cr}}=q_f,p_{\mu
_0}^{cr}=-p_f,t_0^{cr}=-t_f,t^{cr}=-t\right)=q^\mu \left(
q_f,-p_f,-t_f,-t\right)  \label{vb} \\ p_\mu ^{cr}\left( q_0^{\mu
^{cr}}=q_f,p_{\mu
_0}^{cr}=-p_f,t_0^{cr}=-t_f,t^{cr}=-t\right)=p_\mu \left(
q_f,-p_f,-t_f,-t\right)  \nonumber
\end{eqnarray}

These functions do not satisfy the Hamilton equations $\left( \ref{pu}%
\right) .$ These functions do belong to the space of solutions of
the other sheet $p-h=0$, i.e., they are solutions for the other
choice of the direction of time $\left( t\equiv -q=\tau \right) .$
In fact these clock-reversed solutions satisfy the equations
\begin{eqnarray*}
\frac{dq_\mu }{d\left( -q\right) } &=&\frac{\partial h}{\partial
p_\mu } \\ \frac{dp_\mu }{d\left( -q\right) } &=&-\frac{\partial
h}{\partial q_\mu }
\end{eqnarray*}
which are the Hamilton equations corresponding to the choice
\begin{eqnarray*}
q &=&-t \\ p &=&h
\end{eqnarray*}

\subsection{Quantum transformations}

Given a particular solution $\left| \Psi \left( t\right)
\right\rangle $ to the Schr\"{o}dinger equation of a quantum
system its motion-reversed solution (usually called in the
literature ``time-reversed'' solution for the same reasons
mentioned before) is given by
\begin{equation}
\left| \Psi _{mr}\left( t\right) \right\rangle =T\left| \Psi
\left( -t\right) \right\rangle  \label{qw}
\end{equation}
where $T$ is an antiunitary operator which, in coordinate
representation, is equal to the complex conjugation operator
(Ref.\cite {ballentine})
\[
T\Psi \left( q\right) =\Psi ^{\star }\left( q\right)
\]

The transformation $\left( \ref{qw}\right) $ is the quantum
version of the classical motion-reversal transformation $\left(
\ref{oi}\right) $ which means that the transformed solution
$\left| \Psi _{mr}\left( t\right) \right\rangle $ is a solution
for the same Schr\"odinger equation. For example, in the case of a
quantum state $\Psi \left( x,t\right) =e^{-i\left( wt-kx\right) }$
corresponding to a free particle, the transformation $\left(
\ref{qw}\right) $ yields $\Psi _{mr}\left( x,t\right) =e^{-i\left(
wt+kx\right) }$ which corresponds to a state with the same energy,
unfolding in the same direction of time, but with the linear
momentum reversed.

We will now define the quantum version of the classical
clock-reversal transformation $\left( \ref{vb}\right) $ as
\[
\left| \Psi _{cr}\left( t\right) \right\rangle =T\left| \Psi
\left( t\right) \right\rangle
\]
In fact, given a solution $\left| \Psi \left( q\right)
\right\rangle $ of the Schr\"odinger equation
\begin{equation}
i\frac \partial {\partial q}\left| \Psi \left( q\right)
\right\rangle =h\left| \Psi \left( q\right) \right\rangle
\label{xc}
\end{equation}
corresponding to the quantization on the sheet $p+h\left( q_\mu
,p_\mu \right) =0$ $\left( t=q\right) $ with the substitution
$p_i\longrightarrow -i\frac \partial {\partial q_i}$, the time
reversed solution $T\left| \Psi \left( q\right) \right\rangle $ is
not a solution of $\left( \ref{xc}\right) $, but a solution of the
Schr\"odinger equation in the time $t=-q$:
\begin{equation}
-i\frac \partial {\partial q}\left| \Psi \left( q\right)
\right\rangle =h\left| \Psi \left( q\right) \right\rangle
\label{nm}
\end{equation}
corresponding to the quantization on the sheet $p-h\left( q_\mu
,p_\mu \right) =0$. In fact, let us apply the operator $T$ to both
sides of $\left(
\ref{xc}\right) $%
\begin{equation}
-i\frac \partial {\partial q}T\left| \Psi \left( q\right)
\right\rangle =ThT^{-1}T\left| \Psi \left( q\right) \right\rangle
\label{gh}
\end{equation}
Assuming that the reduced Hamiltonian $h$ is real (quadratic in
$p_\mu$) this equation yields
\[
-i\frac \partial {\partial q}T\left| \Psi \left( q\right)
\right\rangle =hT\left| \Psi \left( q\right) \right\rangle
\]
which shows that $T\left| \Psi \left( q\right) \right\rangle $ is
a solution of $\left( \ref{nm}\right) $.

\section{The Wheeler-DeWitt equation}

In the framework of the canonical quantization program the
physical states of the corresponding quantum theory of gravity are
functionals of the spatial metric $g_{ij}$, which satisfy the
quantum version of the classical constraints in accordance with
the Dirac method for quantifying constrained Hamiltonian systems.
The quantization of the momenta constraints implies that the
physical states depends on the geometry $g^3$ of the hypersurfaces
but not on the particular metric tensor $g_{ij}$ used to represent
it. The quantum version of the Hamiltonian constraint is the so
called Wheeler-DeWitt equation $\widehat{H}\Psi=0$.

It was pointed many times the analogy between the Wheeler-DeWitt
equation and the Klein-Gordon equation: both systems have
Hamiltonians which are hyperbolic in the momenta. The space of
solutions of the Klein-Gordon equation can be turned into a
Hilbert space where a subspace with a positive definite inner
product can be defined only if the background is stationary. In
this case the Hilbert space of the physical states will be the
subspace of positive norm, this being equivalent to consider just
one of the sheets of the hyperbolic constraint surface. Beyond
that similarity there is an important difference between both
equations: the Wheeler-DeWitt equation does not have a physical
clock in the configuration space because it does not have a
positive definite potential term to play the role of the mass term
of the Klein-Gordon equation. The double sheet Hamiltonian
structure for the Wheeler-DeWitt equation should therefore be
searched in the phase space. In that case a canonical
transformation should be implemented in order to translate this
double sheet Hamiltonian structure in the phase space of the
original canonical variables to the configuration space of the new
canonical variables. As shown in Section 3 this has been
successfully done for the Taub model (Ref. $\cite{taub}$).
Following the analogy between both equations it was supposed in
Ref. $\cite{taub}$ that each sheet of the Hamiltonian constraint
$p+h=0$ and $p-h=0$ corresponds to positive and negative energies
respectively. In the new approach of this paper each sheet
corresponds to each possible choice in the direction in which the
static trajectory unfolds, being in both cases positive energy
solutions. Actually the stationary solutions to the
Schr\"{o}dinger equation $\left( \ref{xc}\right) $ are
\[
\Psi _{E}\left( t=q,q_{\mu }\right) =e^{-iEq}\varphi \left( q_{\mu
}\right) =e^{-iEt}\varphi \left( q_{\mu }\right)
\]
while the stationary solutions to the Schr\"{o}dinger equation
$\left( \ref {nm}\right) $ are
\[
\Psi _{E}\left( t=-q,q_{\mu }\right) =e^{iEq}\varphi \left( q_{\mu
}\right) =e^{-iEt}\varphi \left( q_{\mu }\right)
\]
which shows that both sets of solutions are positive energy
solutions for the two defined times.

It is important to notice that the presence of a square root
reduced Hamiltonian due to the factorization of the Hamiltonian
constraint leads to a canonical quantization procedure which is
not straightforward. The definition of the operators associated
with this kind of reduced Hamiltonians can be done in two steps:
it is necessary to define the operator under the square root in
order to define, in a second step, the square root itself by means
of the \emph{spectral theorem} (Ref.
\cite{Kuchar},\cite{kuchar2}). This can be done if the operator
under the square root is a positive definite self-adjoint
operator. While this could be done for the Taub model there is no
guarantee that this procedure could be applied to a general case.

\subsection{Boundary conditions for the Wheeler-DeWitt equation}

\smallskip It is a fundamental problem in Quantum Gravity to find proper
boundary conditions in the space of solutions of the
Wheeler-DeWitt equation in order to select the physical solutions.
The Schr\"{o}dinger equation is a parabolic equation while the
original Wheeler-DeWitt equation is an hyperbolic one, having thus
twice the number of solutions than the former. As the Taub model
teaches (Ref. $\cite{taub}$), the connection between the
Schr\"{o}dinger equation and the Wheeler-DeWitt equation is not
straightforward, because the non positive definite potential that
typically appears in the last (see also Ref. $\cite{beluardi}$). A
canonical transformation is necessary in order to find a
Hamiltonian constraint of the form $H=p_0^2-h^2\left( q^\mu ,p_\mu
\right)$ with a well defined reduced Hamiltonian $h$. The system
can thus be quantized by means of the corresponding
Schr\"{o}dinger equation associated with one of the sheets of the
constraint surface (which is equivalent to chose the solutions
associated with the breaking of the clock reversal symmetry). In
Ref. $\cite{taub}$ it was shown that if the proper canonical
transformation to reduce the system is known, it is possible to
provide a criterium to select the physical solutions of the
Wheeler-DeWitt equation. The proposed formalism chooses the
solutions of the Wheeler-DeWitt equation which corresponds to the
solutions of the Schr\"{o}dinger equation for the reduced system.
In order to apply this criterium it is necessary to find a
correspondence between both spaces of solutions. If this
correspondence could be defined it would be possible to transform
the solutions of the Schr\"{o}dinger equation finding in this way
the corresponding physical subspace in the space of solutions of
the Wheeler-DeWitt equation. This amounts to find a quantum
correspondence for relating the wave functions corresponding to a
pair of quantum-mechanical systems whose classical Hamiltonians
are canonically equivalent. In certain cases this quantum
correspondence between both representations can be defined (Ref.
\cite{taub},\cite {gandour}) as
\begin{equation}
\Psi \left( q\right) =N\left( E\right) \int_{-\infty }^{+\infty
}dQe^{iF\left( q,Q\right) }\Phi \left( Q\right)
\end{equation}
where $F\left( q,Q\right)$ is the generating function of the
corresponding canonical transformation. This is a generalization
of the Fourier transformation considered as the quantum version of
the canonical transformation generated by $F\left( q,Q\right)
=qQ.$ Using this quantum correspondence it was possible in Ref.
$\cite{taub}$ to transform the solutions of the Schr\"{o}dinger
equation, so finding the physical solutions of the Wheeler-DeWitt
equation. This procedure amounts to select boundary conditions for
the solutions of the Wheeler-DeWitt equation which are associated
with the direction of time of the chosen physical clock. In this
approach the boundary conditions operates as a symmetry breaking
of the original invariance of the theory under a clock-reversal
transformation. The relation between proper time $T$ and the time
variable $t$ chosen as the physical clock is $dT=Ndt$ where $N$ is
the Lagrange multiplier for the Hamiltonian constraint (or lapse
function). This is a consequence of the way in which the
space-time interval is expressed in the ADM formalism (Ref.
\cite{gravitation}). The proposed approach is thus completely
different from those in which classical proper time is recovered
in a semiclassical regime (Ref. \cite{zeh},\cite {halliwell}). In
these approaches a notion of time is associated with "... the
affine parameter along the histories about which the wave
functions is peaked. So time, and indeed spacetime, are only
derived concepts appropiate to certain regions of configuration
spacetime and contingent upon initial conditions" (Ref. \cite
{halliwell}). On the contrary in the proposed approach there is a
perfectly defined notion of time at the quantum level which plays
the same role as the usual time parameter of the Schr\"{o}dinger
equation in ordinary systems.

The proposed boundary conditions relies on the fact that one knows
how to find the reduced variables in the \emph{classical} level,
i.e., how to separate a physical clock from the whole set of
variables. It would be a great step if one could apply the
underlying physical intuition of this criterium without knowing
how to reduce the system, i.e., without having separated a
physical clock. In fact a fundamental objection against the
reduction formalism applied to canonical quantum gravity is that,
if one considers that quantum mechanics is a more fundamental
theory than classical mechanics, it is not correct to define the
quantum theory using a time variable which was selected by a
classical criterium (the physical clock should be a variable
which monotonically increases along the classical trajectories). In Ref. $%
\cite{taub}$ it was possible to advance in this direction by
proposing a criterium of this kind for time independent reduced
Hamiltonian systems. After performing a change of coordinates the
Hamiltonian constraint could be taken to the form
\begin{equation}
H=p^{2}+V\left( q\right) -h_{q_{\mu }}\left( q_{\mu },p_{\mu
}\right) \label{we}
\end{equation}
The basic fact of the proposed approach is that in the region
where the potential $V\left( q\right) $\ goes to zero the variable
$q$ is the physical clock, i.e., in that region the Hamiltonian
$\left( \ref{we}\right) $ goes to
\begin{equation}
H=p^{2}-h_{q_{\mu }}\left( q_{\mu },p_{\mu }\right)  \label{er}
\end{equation}
\ The boundary conditions to be imposed to the solutions of
Wheeler-DeWitt equation corresponding to the quantization of
$\left( \ref{we}\right) $ is
that its physical solutions should tend, in the region where the potential $%
V\left( q\right) $\ goes to zero, to the solutions of the
Schr\"{o}dinger equation corresponding to $\left( \ref{er}\right)
$, i.e., to functions of
the form $\varphi \left( q,q_{\mu }\right) =\phi \left( q_{\mu }\right) e^{-i%
\sqrt{\varepsilon }q}.$\ In this way, just by analyzing the
asymptotic form of the solutions of the Wheeler-DeWitt equation,
it is possible to select the physical solutions with respect to
the chosen physical clock. This kind of boundary conditions for
the Wheeler-DeWitt equation is similar to those proposed (although
reached by using other methods) in Ref $\cite{wald}.$

In the light of the new interpretative approach, where the choice
of the sheet is equivalent to a symmetry breaking of the
clock-reversal invariance of the theory, this kind of boundary
condition can be reformulated in order to find out the way to
generalize it. The physical meaning of the proposed boundary
conditions is to separate the wave functions going forward in the
time $t=q$ from those going forward in the time $\tilde{t}=-t=-q$.
These two subspaces of solutions of the Wheeler-DeWitt equation
are related to each other by the antiunitary operator $T$
corresponding, in the coordinate representation, to the complex
conjugation operator. The main idea is that the breaking of the
clock-reversal invariance by selecting quantum states which belong
to just one of these subspaces could be a general criterium for
selecting a physical subspace. The real character of the
Wheeler-DeWitt
operator means that, given a solution $\Psi \left( q\right) $, the function $%
\Psi ^{*}\left( q\right) $ is also a solution, these solutions
being linear independent. This means that the space of solutions
of the Wheeler-DeWitt equation $S$ can be decomposed as $S=C\oplus
C^{*}$ where $C$ is a subspace of $S$. A general solution to the
Wheeler-DeWitt equation could be written
as a linear combination of functions belonging to the subspaces $C$ and $%
C^{*}$. It is thus necessary, in order to select the physical
subspace, to decompose the general space of solutions in $C$ and
$C^{*}$. In the case of the Taub model the general solution of the
Wheeler-DeWitt equation can be expressed as a combination of the
modified Bessel functions $K_{\nu }\left( z\right) $ and $I_{\nu
}\left( z\right) $ or as a combination of the modified Bessel
functions $I_{\nu }\left( z\right) $ and $I_{-\nu }\left( z\right)
$. The functions $K_{\nu }\left( z\right) $ and $I_{\nu }\left(
z\right) $ are not related by the clock-reversal operator $T$
while the functions $I_{\nu }\left( z\right) $ and $I_{-\nu
}\left( z\right) $ are, for a real $z$ variable, complex
conjugated of each other: $I_{\nu }\left( z\right) =T\left[
I_{-\nu }\left( z\right) \right] =\left[ I_{-\nu }\left( z\right)
\right] ^{*}$. As a result these two decompositions are not
equivalent, being the decompositions in the basis $\left\{ I_{\nu
}\left( z\right) ,I_{-\nu }\left( z\right) \right\} $ the proper
one in order to select the physical subspace. Then the problem of
imposing boundary conditions on the space of solutions of the
Wheeler-DeWitt equation was reduced to the problem of finding a
proper decomposition of the space of solutions $S$ of the form
$S=C\oplus C^{*}$. These considerations should be complemented by
an analysis of systems with time dependent reduced Hamiltonians.
As it was shown in Ref. $\cite{isham2}$ it is not correct to
quantify these kind of systems by using the unfactored form of the
Wheeler-DeWitt equation because of the very well known ordering of
the quantum operators problem. This case will be address in
another work.

\section{Symmetry breaking of the clock-reversal invariance and the problem
of irreversibility}

Traditionally the main approach for trying to understand the way
in which the phenomenological irreversibility of the world is
compatible with the invariance of the main theories of physics
under ``time-reversal'' (where it is not made the fundamental
distinction between motion-reversal and clock-reversal
transformations) is to separate the whole set of canonical
variables in the relevant ones and those which are irrelevant for
the description of the phenomenological dynamics of the system.
Neglecting this set of irrelevant variables, the set of relevant
ones is an open system whose evolution can not be described as a
Hamiltonian evolution or an unitary evolution in classical or
quantum mechanics respectively. In this way it is possible to
obtain time-asymmetric evolution equations for these open systems
from an initial unstable state condition at time $t=0,$ yielding
e. g. the growing of entropy from $t=0$ towards positive time. But
this can only be considered as a complete solution if we ignore
negative times. In fact for negative times, entropy decreases in a
symmetric way, a fact which reflects nothing but the formation
process of the unstable state (see e. g. \cite{Schulman}).
Therefore the solution is incomplete and must be complemented with
other considerations like those introduced in (Ref.
\cite{casta},\cite{suda},\cite{bohm}). In these works the
emergence of irreversibility depends, in the quantum case, on the
existence of instabilities which forces to use generalized
spectral decompositions of the Hamiltonian with complex
eigenvalues. This generalizations ends in a symmetry breaking of
the ``time-reversal'' invariance of the theory by breaking the
evolution group of the theory in two semigroups. As in the
previous section the space of solutions is decomposed as
$S=C\oplus C^{*}$ where the subspace $C$ is considered as the
space of physical admissible solutions and $C^{*}$ is the space of
the corresponding ``time-reversal'' solutions (the distinction
between motion-reversal and clock-reversal transformations is not
made in those works). This formalism depends on a
particular choice for the space of states of the admissible wave functions $%
\varphi (\omega ,...)$ (where $\omega $ is the energy). In non
relativistic quantum mechanics, several physical considerations
leads to postulate that the space of physical states $C$ is not
the usual space of regular states (Schwarz class wave functions
$S$) but the space of states belonging to the Hardy class from
below $\emph{or}$ from above $H_{\pm }^{2}$ for the variable
$\omega ,$ intersected with $S$ (see Ref.\cite{Gueron}). These
spaces are called $\phi _{-}$ or $\phi _{+\ \ }$ respectively:
\begin{equation}
\phi _{\pm }=\left\{ \left| \psi \right\rangle /\left\langle
\omega _{\mp }\right| \psi \rangle \in \theta \left( S\cap H_{\pm
}^{2}\right) \right\} \label{bn}
\end{equation}
where $S$ denotes the Schwarz class, $H_{\pm }^{2}$ the upper
(lower) Hardy class, and $\theta $ the Heaviside step function.

A complex wave function $f\left( x\right) $ on the real line is a
Hardy class function from above (below) if

1) $f\left( x\right) $ is the boundary value of a function
$f\left( z\right)
$ of a complex variable $z=x+iy$ that is analytical in the half-plane $y>0$ $%
\left( y<0\right) $

2)
\[
\int_{-\infty }^{+\infty }\left| f\left( x+iy\right) \right|
^2dx<k<\infty
\]
for all $y$ such that $0<y<\infty $ $\left( -\infty <y<0\right) $.
The Heaviside step function was introduced in order to have
physical states which vanish for negative energies.

This particular choice of the space of physical states have the
property that each space is not invariant under the action of the
complex conjugation operator (operator $T$ in our formalism) (see
Ref.\cite{casta})
\begin{equation}
T:\phi _{\mp }\rightarrow \phi _{\pm }  \label{zx}
\end{equation}

This means that if one chooses $\phi _{-}$ or $\phi _{+\ \ }$as
the space of physical states then the complex conjugation operator
kicks out any state from the space of admissible functions.
Another fundamental property of Hardy functions is stated in the
Paley-Wiener theorem (Ref.\cite{bohm}). This theorem states that
if $f_{\pm }\left( q\right) \in H_{\pm }$, then the Fourier
transformation
\begin{equation}
g_{\pm }\left( p\right) =\frac 1{\sqrt{2\pi }}\int_{-\infty
}^{+\infty }e^{-ipq}f_{\pm }\left( q\right) dq  \label{ty}
\end{equation}
has the property
\begin{eqnarray}
g_{+}\left( p\right) &=&0 {  for }p<0  \label{rt} \\ g_{-}\left(
p\right) &=&0 {  for }p>0  \nonumber
\end{eqnarray}
This theorem is used to show that the evolution group of the
theory breaks in two semigroups . It can be demonstrated (see Ref
\cite{casta}) that if $\left| \psi \right\rangle \in \phi _{\pm }$
then
\begin{equation}
e^{-iHt}\left| \psi \right\rangle \in \phi _{\pm }{  if} t\
_{>}^{<}\ 0  \label{fg}
\end{equation}
In other words, if $\left| \psi \right\rangle \in \phi _{-}$, then
the evolution operator $U\left( t\right) =$ $e^{-iHt}$ exists in
the physical space $\phi _{-}$ but the inverted time operator
$U\left( t\right) ^{-1}=$ $ e^{iHt}$ does not exist in the space
of physical states $\phi _{-}$, being this the essence of an
irreversible theory.

The underlying intuition of this formalism is similar to the one
which is proposed in the present paper in the sense that it uses a
symmetry-breaking of a ``time-reversal'' invariance which, in
their case,
ends in a time-asymmetric evolution without appealing to any kind of \emph{%
coarse-graining}. Nevertheless if one takes into account the
fundamental distinction made in Section IV between clock-reversal
and motion-reversal it is clear that the formalism proposed in
Ref. \cite{casta} for non-relativistic quantum mechanics must be
adapted to the present relativistic case. The problem is that
non-relativistic quantum mechanics is not invariant under $K$
(clock-reversal transformations) but under $K$ plus $t\rightarrow
-t$ (motion-reversal transformations), which in the old
terminology was known simply as ``time-reversal'' transformation.
In the new terminology classical and quantum mechanics are
invariant under motion-reversal transformations but not under
clock-reversal transformations because, in the context of
parametrized systems formalism, these theories are defined on one
sheet of the Hamiltonian constraint. In the approach of Ref.
\cite{casta} it is supposed that quantum mechanics is invariant
under the operation $K$ and that this invariance is broken by
selecting the space of physical functions as $\phi _{+}$ or $\phi
_{-}$. The new theory defined on these spaces would be no more
invariant under $K$ because of $\left( \ref{zx}\right) $. The
problem is that, taking into account that quantum mechanics is not
invariant under clock-reversal transformations but under
motion-reversal transformations, if we used the old formalism we
would be trying to break a symmetry which is already broken.

Another conceptual problem of this formalism is that its physical
content reduces to the fact that, given a particular unstable
state including the formation process and the decaying process,
the asymmetry is obtained by cutting this process in two halves
corresponding each one to the formation and the decaying process
respectively (this is in fact the breaking of the evolution group
in two semigroups). The whole process is reversible but each part
of it is irreversible. This explanation of irreversibility is
equivalent to the one postulated by Boltzmann after his failure of
deriving irreversibility from the basic laws of mechanics (H
theorem). In a global reversible environment a local unstable
situation like a fluctuation would define two local and opposed
directions of time. In order to explain the global
phenomenological irreversibility it is then necessary to postulate
an initial unstable state without a formation process. The problem
of irreversibility is then taken back to the problem of
postulating certain initial conditions for the universe (see
\cite{Gueron}) and it is not really grounded on a
symmetry-breaking framework.

So we will now propose a way for complementing this previous
approach (Ref.\cite{casta}) in order to set a better grounded
framework for studying the problem of irreversibility based on a
symmetry-breaking of the clock-reversal invariance of General
Relativity. By selecting a sheet of the Hamiltonian constraint one
is not splitting the time line in two halves but unfolding this
whole time line in opposed directions. As it is said in Ref.
\cite{hajicek} by choosing a certain direction of a canonical
variable as time one forces this variable to be non inversible.
This means that its conjugated momentum $p_{t}$ can not change its
sign. In this way, by choosing a physical clock, one forces its
conjugated momentum to have a semi-infinite spectrum. What one
wants to split in two halves is the spectrum of the energy, not
the spectrum of time. One also wants, following the proposed
philosophy, that both sheets of the Hamiltonian constraint
correspond to positive energy solutions. In this way, and roughly
speaking, our proposal is a kind of ``Fourier transformation'' of
the formalism presented in Ref. \cite{casta}. In fact we will
impose the
requirement that the physical states belong to a Hardy class from below $%
\emph{or}$ from above ($\phi _{-}$ or $\phi _{+\ \ }$
respectively) in \emph{time} representation:
\[
\phi _{\pm }=\left\{ \left| \psi \right\rangle /\left\langle
t\right| \psi \rangle \in S\cap H_{\pm }^{2}\right\}
\]

Another way of stating this proposal is saying that a Hardy class
function is, roughly speaking, a ``well behaved'' function whose
Fourier transformation is null in the negative axis. One wants to
have quantum states which, in the energy representation, are null
on the negative axis. In order to satisfy this requirement is
enough to have quantum states which belong to Hardy class
functions in time representation. We will now show that the
requirement $g\left( E\right) \neq 0$ only if $0<E<\infty $ is a
direct consequence of this choice. We will call $\Psi _{\pm }\in
H_{\pm }$ the quantum states corresponding to the choice $t=\mp q$
$\left( p_{t}=\mp p_{q}=\pm h\right) $. Then, using the
Paley-Wiener theorem $\left( \ref{rt}, \ref{ty}\right) $, one
finds
\[
\Phi _{\pm }\left( p_{q}\right) =\frac{1}{\sqrt{2\pi
}}\int_{-\infty }^{+\infty }e^{-ip_{q}q}\Psi _{\pm }\left(
q\right) dq=0 {  if } p_{q} \ \ _{>}^{<}0
\]
which yields
\begin{eqnarray*}
\Phi _{+}\left( p_{q}=-p_{t}=E\right) &=&0 {  if } E<0 \\ \Phi
_{-}\left( p_{q}=p_{t}=-E\right) &=&0 {  if } E<0
\end{eqnarray*}
i.e., for both choices the wave function $G\left( E\right) $ in
the energy representation is zero if $E<0.$ In this way it was
shown that the fact that the physical states vanishes for negative
energies is a direct consequence of the chosen space of admissible
states (Hardy class functions in time representation). It is thus
unnecessary to introduce the Heaviside step function $\left(
\ref{bn}\right) $, being this \ prescription a natural consequence
of the proposed formalism.

\section{Conclusions}

The Hamiltonian constraint of General Relativity is quadratic in
the canonical momenta, which means that the parametrized systems
analogy, with a Hamiltonian constraint linear in the canonical
momentum conjugated to the hidden time, is not completely fitted
to mimic General Relativity. In fact if there is not a privileged
time variable (because of the covariance of the theory under
changes of the foliation of the space-time) all the momenta must
appear in the constraint on an equal foot. Besides, to reduce the
system means to pass from a static trajectory in configuration
space to a trajectory in a reduced configuration space where one
of the original variables was chosen as time. There is no reason
to privilege one or the opposite direction of this variable as
time, which forces the Hamiltonian constraint to be quadratic in
the momentum conjugated to the variable chosen as time (which
could be any monotonically increasing variable along the classical
trajectories). It is then completely necessary to have a
Hamiltonian constraint quadratic in all its momenta circumventing
in this way the objection stated against parametrized systems
approach (Ref. \cite{barbour}) that it cannot explain why the
Hamiltonian constraint of General Relativity is not in fact linear
in one of its momenta (the supposed hidden time). Besides this
conceptual clarification, the proposed interpretation forces to
consider that what changes from one sheet of the constraint to the
other one is not the sign of the energy (as in the Klein-Gordon
analogy) but the direction of time. It was then clearly shown how
both sheets of the constraint corresponds to positive energy
solutions.

General Relativity, differently from ordinary classical or quantum
mechanics, is invariant under the transformation which passes from
one sheet of the Hamiltonian constraint to the other one
(inversion of the direction of time). But it is well known that
ordinary classical or quantum mechanics (theories which, if
parametrized, would be defined on one sheet of the corresponding
Hamiltonian constraint) are also invariant under the so called
``time-reversal'' transformations. In order to solve this apparent
paradox and to show which are the differences between General
Relativity symmetries and ordinary classical or quantum mechanics
symmetries, it was made a distinction between clock-reversals
transformations (a symmetry of General Relativity) and
motion-reversal transformations (a symmetry of General Relativity,
classical and quantum mechanics). This distinction was clearly
formulated both in the classical and quantum levels.

In the light of this new perspective, the boundary conditions on
the space of solutions of the Wheeler-DeWitt equation proposed in
Ref. $\cite{taub}$ were restated. The main idea is that the
boundary conditions to be imposed should act as a
symmetry-breaking of the clock-reversal symmetry of General
Relativity. In the quantum level the Wigner operator $T$ (complex
conjugation in coordinate representation) is the operator which
transforms quantum states defined on one sheet of the constraint
to the other one. We can thus say that the problem of finding
proper boundary conditions on the space of solutions $S$ of the
Wheeler-DeWitt equation for time independent systems was reduced
to the problem of finding a realization of that space of the form
\begin{equation}
S=C\oplus C^{*}  \label{cv}
\end{equation}
where the passage from one subspace $C$ to the other one $C^{*}$
is generated by $T.$ If this decomposition can be done (which is
always possible for systems with time independent reduced
Hamiltonians), one can break the clock-reversal symmetry by
defining the space of physical solutions as one of this subspaces.
One obtains a theory which is no more invariant under
clock-reversals transformations but which still has the
motion-reversal symmetry (as usual classical or quantum
mechanics). In this way it is possible to define boundary
conditions for Wheeler-DeWitt equation with the definite meaning
of selecting one conventional direction of time with respect to a
chosen physical clock.

In the context of finding a formalism for understanding
irreversible process in a complementary way with the
\emph{coarse-graining} approaches, it was proposed in previous
papers (Ref.\cite{casta}) a similar decomposition of the space of
states of the quantum systems under study of the form $\left(
\ref{cv}\right) ,$ using the Hardy class functions from above and
below. We review this approach, state its main problems and
suggest an adaptation of it in order to satisfy the requirements
of the proposed interpretation. The main idea is to consider that
the subspace of physical states is the subspace of quantum states
which are Hardy class functions from above or below in the time
representation (and not in the energy representation as in the
previous approach, which intended to break the symmetry under
motion reversal transformations). We think that, by doing this, we
set a much more stronger and clearer ground for studying the
emergence of irreversibility taking into account the symmetries of
General Relativity under clock-reversal transformations.

\section{Appendix: Comment on the time operator.}

It could be argued that if one selects a physical clock for
measuring time nothing prevents us from considering the chosen
dynamical variable as a quantum variable. The rough quantum
mechanical distinction between dynamical variables associated with
operators and time, which is supposed to be just an evolution
parameter, should disappear if one considers that there is not
something as {\bf Time} but only dynamical variables playing the
role of physical clocks. If one assume that this distinction
should not hold any more in the context of an operational
definition of time then it should be possible to circumvent the
well known impossibility of associating a quantum operator with
time, which is said to be a result of the fact that the
Hamiltonian is semi-bounded from below. This result was obtained
by Pauli in 1933. Briefly the statement that the time operator
$T$\ does not exist if the spectrum of $H$\ is bounded from below
(Ref.\cite{ballentine}). If this operator could be defined then a
state $\left| E\right\rangle $\ could be transformed in a state of
any energy $E+\alpha $\ with arbitrary real $\alpha $\ by applying
the unitary operator $e^{i\alpha T}\left| E\right\rangle .$\ In
fact the energy of this transformed state is
\begin{equation}
He^{i\alpha T}\left| E\right\rangle =\left( E+\alpha \right)
e^{i\alpha T}\left| E\right\rangle  \label{dc}
\end{equation}
which is inconsistent with the assumption that the spectrum of
$H$\ is bounded from below. But, as was stated above, the
necessity of defining a time operator is a main problem if one
follows the proposed interpretation of Canonical Quantum Gravity.
The covariance of General Relativity forbids us to consider a
certain variable as a privileged clock and consider it, in the
quantum version of the theory, as a c-number. This fact imposes
the necessity of facing the problem of defining a time operator.
As was said above by choosing a physical clock, one forces its
conjugated momentum to have a semi-infinite spectrum. Then, when
one breaks the clock-reversal symmetry by choosing a direction for
time, the possibility of associating an operator with time is
apparently eliminated. But it is known that in fact exist several
examples of self-adjoint operators that do not possess spectra
spanning the entire real line, e.g., the momentum and position
operators of a particle trapped in a box, the angular momentum and
the angle operators, the harmonic oscillator number and phase
operators (Ref.\cite {galapon}) . Following this examples it was
shown in Ref. \cite {galapon} that a time operator conjugated to a
Hamiltonian with a semi-bounded spectrum can be consistently
defined. The fact that the spectrum of the reduced Hamiltonian is
bounded from below is not more problematic, for a quantum
definition of the corresponding operators, than the case of a
particle constrained by certain boundary conditions to move in the
positive real semi-axis. The main difference between these cases
is that in ordinary quantum mechanics it is enough to consider
time as a classical external parameter of evolution, being
unnecessary to associate it with an operator. It is only in the
context of the quantization of gravity that a physical motivation
for defining a time operator appears, because in General
Relativity, and according to the proposed formalism, there is not
an external classical time, but only dynamical variables which
could play the role of physical clocks. In a quantum context it
would be inconsistent to treat the physical clock as a classical
parameter while the rest of the dynamical variables are associated
with quantum operators, because there is not an essential
difference between the variable used as a physical clock and the
rest of the canonical variables, being this a consequence of the
covariance of the theory. If this were the correct physical
interpretation, the proposed approach imposes the necessity of
using the formal demonstration stated in Ref. \cite{galapon} for
defining a time operator.

The fact that it is forbidden to use transformations like $\left(
\ref{dc} \right) $\ is a common feature of any symmetry breaking
formalism. The whole space of solutions of the Wheeler-DeWitt
equation has a symmetry generated by the unitary operator
$e^{i\alpha T}$. The breaking of this symmetry by imposing the
proposed boundary conditions on this space of solutions defines
two subspaces which are no more invariant under the group
generated by $ e^{i\alpha T}$. Certainly this breaking do not
yields the non existence of $ T $. In fact, if one has quantum
states which belong to a Hardy class functions in time
representation, the relations $\left( \ref{fg},\ref{gh} \right) $\
change to the assertion that if $\left| E\right\rangle _{\pm }\in
H_{\pm }$\ then
\[
e^{-iT\alpha }\left| E\right\rangle _{\pm }\in H_{\pm } {  if }
\alpha >0
\]
In this way the Pauli theorem can be circumvented by choosing the
space of admissible quantum states to be a Hardy class function
from above or below, i.e., by an explicit breaking of the
clock-reversal symmetry of the space of solutions of the
Wheeler-DeWitt equation. After choosing a direction for the
evolution of the physical clock, the space of admissible states
for $g\left( E\right) $\ is restricted to those functions $g\left(
E\right) $\ such that $ g\left( E\right) \neq 0$\ if $0<E<\infty
$.

\bigskip

{\bf Acknowledgments}

This work was supported by Consejo Nacional de Investigaciones
Cient\'\i ficas y T\'ecnicas, Universidad de Buenos Aires (Proj.
X-143) and Fundación Antorchas.

\newpage
Figure 1: (a) represents a particular motion in which the variable
$q_{1}$ was chosen as the physical clock $t$ ($t=q_{1}$), (b)
represents the motion reversal of (a) ($q_{1}$ is still the time
$t$), (c) represents the clock-reversal of (a) (the time $t$ is
now equal to $-q_{1}$) and (d) represents the clock-reversal of
(b) or the motion reversal of (c).

\end{document}